\documentclass[review]{elsarticle}

\usepackage{lineno,hyperref}
\usepackage{amsmath}
\usepackage{graphicx,psfrag,color}
\usepackage{dsfont}

\modulolinenumbers[5]

\journal{International Journal of Engineering Science}

\newcommand{\dsx}{\mathtt{x}}
\newcommand{\dsy}{\mathtt{y}}
\newcommand{\dsz}{\mathtt{z}}
\newcommand{\dsu}{\mathtt{u}}

\begin{document}

\begin{frontmatter}

\title{ Anti-plane surface waves in media with surface structure: discrete vs. continuum model 
}

\author[PG,sfedu]{Victor A.\ Eremeyev\corref{cor1}}
 \ead{eremeyev.victor@gmail.com}
 \author[IITK]{Basant Lal Sharma }
 \ead{bls@iitk.ac.in}

 \address[PG]{Faculty of Civil and
 Environmental Engineering, Gda\'{n}sk University of Technology, \\ ul. Gabriela Narutowicza 11/12
 80-233 Gda\'{n}sk, Poland}
\address[sfedu]{Don State Technical University, Gagarina sq., 1, 344000 Rostov on Don, Russia}
\address[IITK]{Department of Mechanical Engineering, Indian Institute of Technology Kanpur, Kanpur 208016, India}
\cortext[cor1]{Corresponding author. Tel.: +48 58 3471891 }
%
%
%

\begin{abstract}
We present a comparison of the dispersion relations derived for anti-plane surface waves using the two distinct approaches of the surface elasticity vis-a-vis the lattice dynamics. We consider an elastic half-space with surface stresses described within the Gurtin--Murdoch model, and present a formulation of its discrete counterpart that is a square lattice half-plane with surface row of particles having mass and elastic bonds different from the ones in the bulk. As both models possess anti-plane surface waves we discuss similarities between the continuum and discrete viewpoint. In particular, in the context of the behaviour of phase velocity, we discuss the possible characterization of the surface shear modulus through the parameters involved in lattice formulation.
\end{abstract}


\end{frontmatter}

\linenumbers

\section{Introduction}
 Recent advances in nanotechnology have resulted in growing interest to the application of discrete and continuum models for the description and understanding of the phenomena at the nanoscale. In particular, the surface elasticity model proposed by \cite{GurtinMurdoch1975a,gurtin1978surface} and its further extensions have found significant applications to the modelling of material behaviour at the nanoscale, see, e.g., \cite{Duanetal2008,Wang2011,javili2013thermomechanics,eremeyev2016effective}. Indeed, surface elasticity has been found very useful in the description of such phenomena due to a prominent size-effect observed for the nano-structured materials. As in the general framework of continuum mechanics, a key problem of the material description, also within the Gurtin--Murdoch model, is the determination of the additional material parameters such as surface elastic moduli and surface mass density. A straightforward experimental approach to their measurement has been presented by \cite{cuenot2004surface,jing2006surface,xu2017direct} but it requires rather complex techniques as well as some additional assumptions concerning the material behaviour and the used model. An alternative approach uses the numerical technique of molecular dynamics simulations, see, e.g., \cite{miller2000size,shenoy2005atomistic}, where the surface elastic moduli are determined from direct atomistic simulations. Let us note that the lattice dynamics as described by \cite{brillouin1946wave,BornHuang1985} provides the possibility to solve various dynamical problems involving waves with attention focussed on the influence of microstructure in as much detail as possible, including even the surface microstructure, see \cite{slepyan2012models,mishuris2007waves,mishuris2009localised,porubov2013nonlinear,sharma2017scattering,porubov2018two} and the references therein.

The aim of this paper is to characterize the material parameters used in the linear Gurtin--Murdoch model through the lattice model parameters. To this end we consider the anti-plane surface waves and compare the dispersion relations derived within the continuum and discrete model.

The paper is organized as follows. First, following \cite{eremeyev2016surface} in Section \ref{GMmodel}, we briefly review the anti-plane surface waves in an elastic half-space assuming the presence of surface stresses within the linear Gurtin--Murdoch model. The dispersion relation is derived. In Section~\ref{LDmodel} using the technique described by \cite{sharma2015diffraction,sharma2015diffraction2,sharma2017linear} we find dispersion relation for a square lattice with one surface row of particles which masses and bonds stiffness are different from others in the bulk. Finally, in Section~\ref{GMLDcomparison} we discuss similarities between the continuum and the discrete model.

\section{Gurtin--Murdoch model of surface elasticity}\label{GMmodel}
\begin{figure}[!h]
\psfrag{A}[m][][0.8]{$1)$} \psfrag{B}[m][][0.8]{$2)$}
 \psfrag{a}[m][][0.8]{$a$} \psfrag{k}[m][][0.8]{$\alpha K$}\psfrag{N}[m][][0.8]{$M$}\psfrag{K}[m][][0.8]{$K$}
 \psfrag{n}[m][][0.8]{$mM$}
 \psfrag{c}[m][][0.71]{$\dsy=0$}
 \psfrag{d}[m][][0.7]{$\dsy=-1$}
 \psfrag{e}[m][][0.7]{$\dsy=-2$}
 \psfrag{x}[m][][1]{$x$}
 \psfrag{y}[m][][1]{$y$}
 \psfrag{z}[m][][1]{$z$}
  \psfrag{f}[m][][1]{$\dsx$}
 \psfrag{g}[m][][1]{$\dsy$}
 \psfrag{h}[m][][1]{$\dsz$}
 \psfrag{X}[m][][1]{$\mathbf{i}$}
 \psfrag{Y}[m][][1]{$\mathbf{j}$}
 \psfrag{Z}[m][][1]{$\mathbf{k}$}
 \psfrag{v}[m][b][0.7]{$\dsx=-1$}
 \psfrag{u}[m][][0.7]{$\dsx=0$}
 \psfrag{w}[m][t][0.7]{$\dsx=1$}
 \centering\includegraphics[width=5.2in]{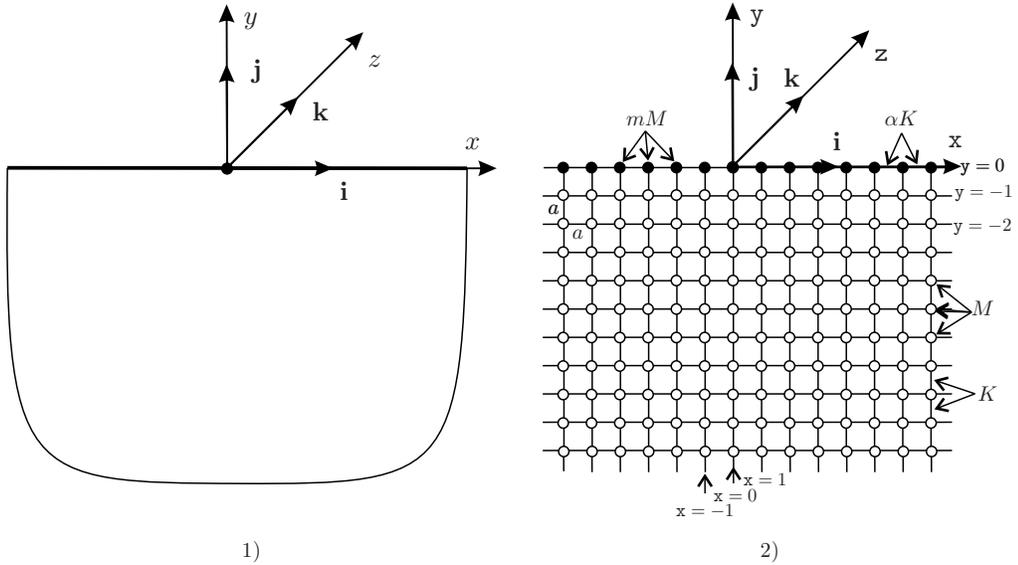}
\caption{1) Half-space with surface stresses and 2) a square lattice with different particles at the surface.}
\label{Halfspace}
\end{figure}

Let us consider a three-dimensional elastic half-space $y\le 0$, where $x$, $y$, $z $ are Cartesian coordinates, and $\mathbf{i}$, $\mathbf{j}$, $\mathbf{k}$ are the unit basis vectors, see Fig.~\ref{Halfspace}(1). On the free boundary $y=0$ we assume the action of surface stresses described within the model of surface elasticity by \cite{GurtinMurdoch1975a,gurtin1978surface}.
Within the linear Gurtin-Murdoch model it has been shown earlier that anti-plane surface waves can be constructed, see \cite{eremeyev2016surface}.
For anti-plane deformation, given by the displacement $\mathbf{u}(x,y,t)=u(x,y,t)\mathbf{k}$, the equation of motion for $x\in\mathds{R}, y<0$ is, see e.g. \cite{achenbach2012wave},
\begin{equation}\label{conteqmotion}
\mu\left(\frac{\partial^2}{\partial x^2} +\frac{\partial^2}{\partial y^2}\right)u(x,y,t)=\rho \ddot{u}(x,y,t),
\end{equation}
while for the boundary $y=0$ we have
\begin{equation}\label{bccondition}
\mu \frac{\partial }{\partial y}u(x,y,t)=\mu_s\frac{\partial^2}{\partial x^2}u(x,y,t)-\rho_s \ddot{u}(x,y,t),
\end{equation}
where $\mu$ is the shear modulus, $\rho$ is the mass density, $\mu_s$ and $\rho_s$ are the surface shear modulus and mass density, respectively.
For infinitesimal anti-plane motions within the Gurtin--Murdoch model, given $\mu$ and $\rho$ for the bulk, one also needs to find the two surface parameters that is $\mu_s$ and $\rho_s$.

Considering an anti-plane surface wave of the form
\begin{equation}\label{defwaveC}
u(x,y,t)=u_0\exp({ikx-i\omega t})\exp({\gamma y}),
\end{equation}
where $\omega$ is the circular frequency, $k$ is the wave number, $u_0$ is the amplitude, and $i$ is the imaginary unit, we find the dispersion relation
\begin{equation}\label{dispersionequationC}
 \mu\gamma(k,\omega)= \rho_s\omega^2 - { \mu_s k^2},\quad \gamma=\gamma(k,\omega)\equiv \left(k^2 - \frac{\omega^2}{c_T^2} \right)^{1/2},
\end{equation}
where $c_T=\sqrt{\mu/\rho}$ is the shear wave speed,
see \cite{eremeyev2016surface} for details.
Introducing the phase velocity $c=\omega/k$ and the characteristic length $r=\rho_s/\rho$, we transform \eqref{dispersionequationC} into the dimensionless form
\begin{equation}\label{DispEqHS333}
 \frac{c^2}{c_T^2} =\frac{ c_s^2 }{c_T^2} +\frac{1}{|k|r} \sqrt{1-\frac{c^2}{c_T^2}},
\end{equation}
where $c_s=\sqrt{{\mu}_s/\rho_s}$ is the shear wave speed in the thin film associated with the Gurtin--Murdoch model. The solution of \eqref{DispEqHS333} exists if and only if $c$ is in the range $c_s<c\le c_T$. In addition, we have the following properties $c(0)=c_T$ and $c(k)\to c_s$ at $k\to\infty$.

\section{Lattice Model}\label{LDmodel}
Following the technique by \cite{brillouin1946wave,mishuris2009localised,sharma2015diffraction,sharma2017linear} let us consider a square lattice which occupies half-space $\dsy\le 0$ as shown in Fig.~\ref{Halfspace}(2). The positions of the lattice particles are described through its lattice
coordinates $\dsx\in \mathds{Z}$, $\dsy\le0$, $\dsy\in \mathds{Z}$. The lattice mostly consists of identical particles of mass $M$ connected to each other by linearly elastic bonds (springs) of stiffness $K$. In order to model surface tension we assume that the free surface $\dsy=0$ is constituted by particles with masses $m M$ and bonds with spring constant $\alpha K$, whereas $m$ and $\alpha$ are dimensionless parameters. The anti-plane displacement of a particle, indexed by its lattice
coordinates $\dsx\in \mathds{Z}$, $\dsy\in \mathds{Z}$, is denoted by $\dsu_{\dsx,\dsy}$. Herein and after, let $\mathds{Z}$ denote the set of integers.
The motion equation for square lattice is given by
\begin{equation}\label{eqmotion}
 M \ddot{\dsu}_{\dsx,\dsy}=K\left({\dsu}_{\dsx+1,\dsy}+{\dsu}_{\dsx-1,\dsy}+{\dsu}_{\dsx,\dsy+1}+{\dsu}_{\dsx,\dsy-1}-4{\dsu}_{\dsx,\dsy}\right)
\end{equation}
for $\dsx\in {\mathds{Z}}$, $\dsy<0$, $\dsy\in {\mathds{Z}}$, see, e.g. \cite{sharma2017linear}. On the free surface that is for $\dsx\in {\mathds{Z}}$, $\dsy=0$ we have
\begin{equation}\label{eqmotionBC}
 mM \ddot{\dsu}_{\dsx,\dsy}=\alpha K\left({\dsu}_{\dsx+1,\dsy}+{\dsu}_{\dsx-1,\dsy}-2{\dsu}_{\dsx,\dsy}\right)+K\left({\dsu}_{\dsx,\dsy-1}-{\dsu}_{\dsx,\dsy}\right).
\end{equation}
Let us consider the discrete analogue of the surface wave form \eqref{defwaveC}, i.e.,
\begin{equation}\label{defwave}
{\dsu}_{\dsx,\dsy}={\dsu}_0\exp({i\xi \dsx-i\omega t})\exp({\eta \dsy}),
\end{equation}
where $\xi$ is the discrete wave number, $\xi\in(-\pi,\pi)$, and $\eta$ is assumed to be positive.
It is found that $\omega$ and $\eta$ satisfy the two equations
\begin{align}\label{eq1}
 &-M\omega^2=K\left(2\cos\xi+2\cosh\eta-4\right),
 \\\label{eq2}&-mM\omega^2=\alpha K\left(2\cos\xi-2\right)+K (\exp(-\eta)-1).
\end{align}

Motivated by a continuum context \citep{sharma2015diffraction}, let
\begin{equation}\label{defMKcTcS}
M=\rho a^3,\quad K=\mu a,
\end{equation}
where $\rho$ and $\mu$ are the mass density and shear modulus introduced in Section~\ref{GMmodel}.
Then from \eqref{eq1} and \eqref{eq2} we get
\begin{align}\label{eq3}
 \omega^2=&\frac{c_T^2}{a^2}\left(4-2\cos\xi-2\cosh\eta\right),\\
 \label{eq4}
 \omega^2=&2\frac{\alpha c_T^2}{m a^2}\left(1-\cos\xi\right)+ \frac{c_T^2}{m a^2}(1-\exp(-\eta)).
\end{align}
These two equations result in a dispersion relation for the surface waves on square lattice half-plane with surface structure. As $\xi$ and $\eta$ play a role of $k$ and $\gamma$, respectively, Eqs. \eqref{eq3} and \eqref{eq4} are the discrete analogues of the dispersion relation \eqref{dispersionequationC} for the elastic half-space with surface stresses.

\section{Comparison of surface wave dispersion}\label{GMLDcomparison}
\begin{figure}[!h]
  \psfrag{a}[m][][1]{$c_T$} \psfrag{b}[m][][1]{$c_s$}
  \psfrag{k}[m][][1]{$k$}\psfrag{g}[m][][1]{$c_{GM}$}\psfrag{d}[m][][1]{$c_o$}
  \psfrag{l}[m][][1]{$c_{lm}$}
 \psfrag{c}[m][][1]{$c$}
  \psfrag{p}[m][][1]{$\pi/a$}
  \psfrag{A}[l][][0.7]{$c=c_{GM}(k)$ { is the phase velocity  for the Gurtin--Murdoch model given by \eqref{DispEqHS333}}}
  \psfrag{B}[l][][0.7]{{$c=c_{lm}(k)$ is the phase velocity  for the lattice model given by \eqref{eq3} and \eqref{eq4}}}
  \psfrag{C}[l][][0.7]{{$c=c_o(k)$ is the phase velocity  for an infinite square lattice given by \eqref{ck-lattice}}}
 \centering\includegraphics[width=4.2in]{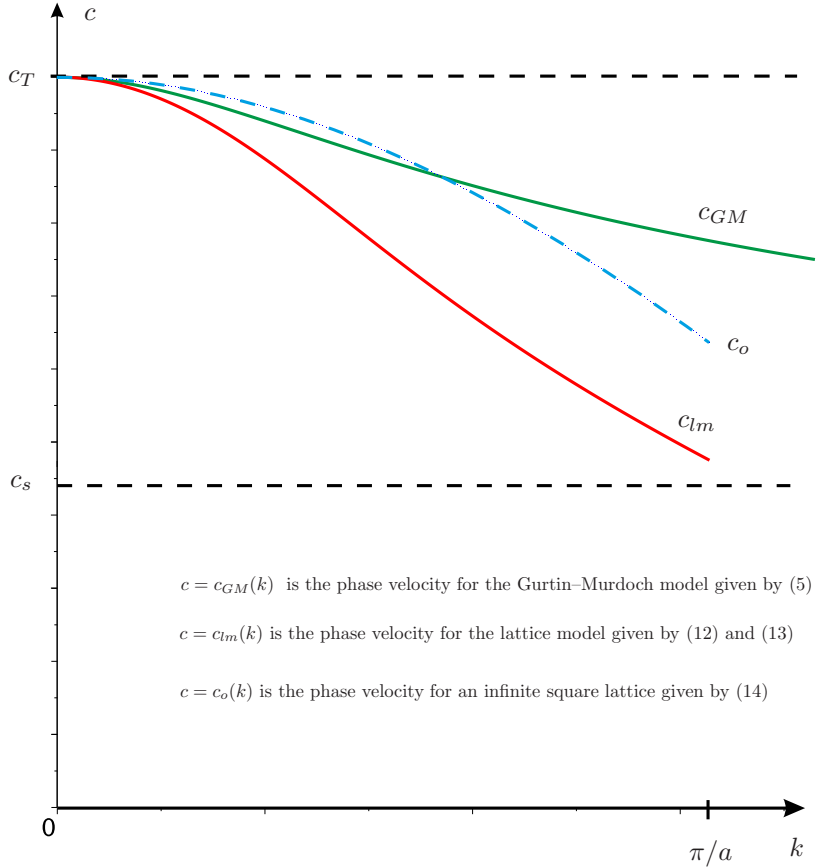}
\caption{Phase velocity vs. wave number for discrete and continuum model.}
\label{Disersionrelations}
\end{figure}

In order to compare the dispersion relation \eqref{DispEqHS333} with \eqref{eq3} and \eqref{eq4} we substitute $\xi=k a$ and consider $k$ in the range $k\in [0,\pi/a]$. The typical dispersion curves is shown in Fig.~\ref{Disersionrelations}. All curves   start from the point $(0, c_T $) with a
horizontal tangent $c=c_T$. Two horizonal dashed lines correspond to $c=c_T$ and $c=c_s$, respectively. Curves $c=c_{GM}(k)$ and $c=c_{lm}(k)$ present the solutions of \eqref{DispEqHS333} for the Gurtin--Murdoch model and  of \eqref{eq3} and \eqref{eq4} for the lattice model.
The dashed blue curve in Fig. \ref{Disersionrelations} corresponds to the equation
\begin{equation}\label{ck-lattice}
 c=c_o(k)\equiv 2 c_T \frac{\left|\sin\left(\frac{ka}{2}\right)\right|}{ka},
\end{equation}
which gives the phase velocity $c_o$ for an infinite square lattice \citep{brillouin1946wave,sharma2017linear}. Here we used the following values of material parameters: $c_T=1$, $c_s=\sqrt{0.2}$, $r=0.005$ for continuum model and $a=0.01$, $\alpha=0.1$ and $m=0.5$ for the lattice. Note that these parameters are chosen to satisfy the relation
\begin{equation}\label{normalization1}
    c_{GM}(0)=c_T=c_{lm}(0),
\end{equation}
which constitutes the first correspondence between continuum and discrete model. From \eqref{normalization1} we get the relation
\[c_T =\sqrt{\frac{\mu}{\rho}}=a\sqrt{\frac{K}{M}},\]
which is consistent with assumption \eqref{defMKcTcS}. So for long wave approximation ($k\approx 0$) we have good coincidence between both discrete and continuum model.

Clearly, while keeping $m$ and $\alpha$ constant as $a\to 0$ we cannot obtain anti-plane surface wave as a continuum limit of the discrete model, since in this case we recover an elastic half-space for which it is well known that such waves {\em do not exist} \citep{achenbach2012wave}. Hence, to capture the behaviour of the Gurtin--Murdoch model one needs to apply an appropriate scaling for $m$ and $\alpha$.

\begin{figure}[!h]
  \psfrag{a}[m][][1]{$c_T$} \psfrag{b}[m][][1]{$c_s$}
  \psfrag{k}[m][][1]{$k$}\psfrag{g}[m][][1]{$c_{GM}$}\psfrag{d}[m][][1]{$c_{lm}$}
  \psfrag{l}[m][][1]{$c_{o}$}
 \psfrag{c}[m][][1]{$c$}
  \psfrag{p}[m][][1]{$\pi/a$}
 \centering\includegraphics[width=5.2in]{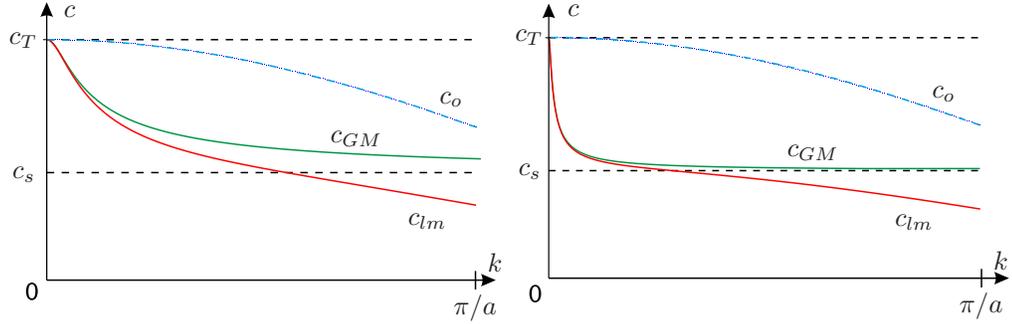}
\caption{Phase velocity vs. wave number for discrete and continuum model for $a=0.001$ (on the left) and $a=0.0001$ (on the right).}
\label{Disersionrelations2}
\end{figure}

We propose the following scaling law
\begin{equation}\label{scalinglaw}
    \alpha=\frac{1}{a}\frac{\mu_s}{\mu},\quad m=\frac{1}{a}\frac{\rho_s}{\rho}.
\end{equation}
With \eqref{scalinglaw} we find that the surface bond stiffness becomes constant as $a\to 0$, $\alpha K={\mu_s} $, whereas the mass  of surface particles $mM=\rho_s a^2$. As a results, for $c_s$ we have
\begin{equation}\label{normalization2}
    c_s=\sqrt{\frac{\mu_s}{\rho_s}}=\sqrt{\frac{\alpha K}{m M}} \, a=\sqrt{\frac{\alpha  }{m  }} \, c_T.
\end{equation}
Thus, the scaling law \eqref{scalinglaw} establishes the second correspondence between continuum and discrete model or, more precisely, between lattice model with surface particles different from the ones in the bulk and the Gurtin--Murdoch model of surface elasticity. Using \eqref{scalinglaw} in Fig.~\ref{Disersionrelations2} we present the dispersion relations for $a=0.001$ and $a=0.0001$. In general, the choice of scaling laws is not unique, one may propose another one as well.

Let us note that the relations between the linear Gurtin--Murdoch model and the lattice model in a certain sense similar to relations between surface elasticity and the Toupin--Mindlin linear strain gradient elasticity. Indeed, both theories possess surface energy and the corresponding dispersion relations for anti-plane surface waves are qualitatively similar for both models \citep{eremeyev2018comparison}. The relations between material parameters of these models can be obtained from the equations
\[c_{GM}(0)=c_T=c_{TM}(0),\quad \lim\limits_{k\to\infty}c_{GM}(k)=c_s=\lim\limits_{k\to\infty}c_{TM}(k),\]
where $c_{TM}=c_{TM}(k)$ is the phase velocity for the Toupin--Mindlin constitutive relations.
Nevertheless, there is difference in decay with the depth, so their correspondence is not straightforward as in presented case here. In addition, for the discrete model $c_{lm}(k)$ is defined for the finite range of $k$.

\section*{Conclusions}
For anti-plane surface waves, we demonstrate the essential similarity between dispersion relations derived within both discrete and continuum model of a surface structure. We consider a square semi-infinite lattice with a surface row of particles which properties are different from ones in the bulk, and the linear Gurtin--Murdoch model of surface elasticity. These different models can capture material behavior related to presence of surface energy. On the other hand the transition from the lattice model to the Gurtin--Murdoch model is not straightforward, as it requires additional assumptions on the dependence of surface particles' mass and surface bond stiffness on the lattice cell length $a$.

\section*{Acknowledgments} V.A.E. acknowledges the support of the Government of the Russian Federation (contract No. 14.Z50.31.0046).
B.L.S. acknowledges the support of SERB MATRICS grant MTR/2017/000013.

\section*{References}
\bibliographystyle{model2-names}\biboptions{authoryear}
\bibliography{latticeantiplane}

\end{document}